\documentclass[doublecol,figures]{epl2} 

\title{Variational approximations for stationary states of Ising--like models}
\shorttitle{Variational approximations to stationary states} 

\author{A. Pelizzola\inst{1,2,3}}
\shortauthor{A. Pelizzola}

\institute{                    
  \inst{1} Dipartimento di Scienza Applicata e Tecnologia, CNISM and
  Center for Computational Studies, Politecnico di Torino,
Corso Duca degli Abruzzi 24, I--10129 Torino, Italy\\
  \inst{2} INFN, Sezione di Torino, via Pietro Giuria 1, I-10125 Torino,
Italy\\
\inst{3} Human Genetics Foundation, HuGeF, Via Nizza 52, I-10126 Torino,
Italy
}

\abstract{
We introduce a new variational approach to the stationary state of
kinetic Ising--like models. The approach is based on the cluster
expansion of the entropy term appearing in a functional which is
minimized by the system history. We rederive a known mean--field
theory and propose a new method, here called diamond approximation,
which turns out to be more accurate and faster than other methods of
comparable computational complexity. 
}

\begin{document}

\maketitle

\section{Introduction}

In equilibrium statistical mechanics, the exact solution of a model
with a large number of interacting variables is most often
analytically unfeasible and computationally intractable. For this
reason, many approximate methods have been developed in the last
century to deal with this problem. Among these, mean--field--like
techniques play a fundamental role and are still the subject of a
research activity aimed to improve accuracy and speed and to refine
theoretical foundations. The importance of these techniques can be
understood by considering that they are simple tools, easily
applicable to many different models, and usually much faster than
Monte Carlo simulations. Moreover, they can provide exact results in
special cases. 

Nonequilibrium statistical mechanics, although not as well developed
as its equilibrium counterpart, poses similar problems and
mean--field--like techniques have been and are being adapted for
application to out of equilibrium models. In this context, a typical
and well defined problem, which actually goes far beyond the
boundaries of statistical mechanics, is finding the stationary state
of a Markov process \cite{vanKampen}. 

In developing mean--field--like techniques for this problem,
statistical physicists often focused on specific examples, like
kinetic Ising and Ising--like models, or epidemic processes. In trying
to go beyond the simple mean--field theory \cite{Kawasaki}, several
approaches were developed. Here we cannot make an exhaustive review,
but we shall try to briefly recall the approaches which are the most
relevant for the present proposal.

One line of approach, which sometimes goes under the name of local
equilibrium approximation \cite{Kawasaki,AdvPhys,DeLos1,DeLos2}, is
based on the assumption that in the stationary state, the probability
distribution of a model with many variables factors into a suitable
product of marginals, each involving a small number of variables. In
the extreme case, when one assumes factoring into a product of
single--variable marginals, a mean--field theory is obtained. 

Another technique is the path probability method (PPM)
\cite{PPM0,PPM1,PPM2,PPM3} (the dynamical version of the cluster
variation method \cite{CVM,An,MyRev}), where the kinetic problem
is written in terms of a 2--times variational problem, and the entropy
terms appearing in the kinetic functional to be minimized are
approximated by means of a cluster expansion. This technique is not
unrelated to the local equilibrium approximation, and in specific cases
equivalence has been rigorously proved \cite{ZM-JSTAT}. 

In addition, a very recent proposal is the so--called dynamic cavity
method \cite{DynCavity00,DynCavity0,DynCavity1,DynCavity2,DynCavity3},
an interesting generalization of the well--known and widely applied
cavity method \cite{Mezard} to kinetic problems. This is a
message--passing algorithm, which has been shown to be efficient on
large systems and to perform better than mean--field theories
\cite{DynCavity3}.

In the present letter we try to make a step forward in this line of
research by proposing a new variational method for the stationary
state problem. We retain the cluster expansion idea on which the PPM
is based, but instead of applying it to a 2--times functional, we
start with the functional which describes the full system history. In
this way, we have more freedom in the choice of the clusters which
enter the expansion. While in PPM the main clusters are obtained by
selecting a suitable set of interacting variables and using the same
variables for two consecutive times, our main clusters can involve
more than 2 consecutive times and need not be time invariant, that
is different variables can be involved at different times. We
illustrate our idea by two examples: the first one, termed star
approximation, reduces to an already known mean--field theory; in the
second one, termed diamond approximation, the choice of the main
clusters is inspired by the dynamic cavity method. The result is a
new method, more general than dynamic cavity, which, compared with
other techniques of comparable complexity, will prove to be more
accurate and faster.

\section{Variational approximations}

We consider a model with discrete variables $s_i$, associated to the
nodes $i = 1, 2, \cdots N$ of an undirected graph $G = (V,E)$, where
$V$ denotes the set of nodes and $E$ the set of edges. The
neighbourhood of a node $i$ is denoted by $\partial i = \{ j \in V :
(i,j) \in E \}$, and its cardinality $d_i = \vert \partial i \vert$ is
called the degree of node $i$.

The variables $s_i$ take values in a finite set, which is usually
(but does not need to) the same for all nodes $i$. Typical examples
are the Ising model, where $s_i \in \{ +1, -1 \}$, the Potts model,
where $s_i \in \{ 0, 1, \cdots q \}$, epidemic models, where $s_i \in
\{\tx{Susceptible}, \tx{Infected}, \tx{Recovered} \}$ or variants
thereof. The value of $s_i$ at time $t$ is denoted by $s_i^t$, and the
state of the system at time $t$ by $s^t = \{ s_1^t, s_2^t, \cdots s_N^t
\}$. The kinetics we consider is formulated in terms of a
discrete time Markov process,
\begin{equation}
\label{eq:Markov}
P(s^{t+1}) = \sum_{s^t} P(s^t) W(s^t \to s^{t+1}),
\end{equation}
specified through the transition matrix
\begin{equation}
\label{eq:W}
W(s^t \to s^{t+1}) = P(s^{t+1} \vert s^t),
\end{equation}
that is the conditional probability of the state at time $t+1$, given
the state at time $t$. The transition matrix must obey the
normalization condition 
\begin{equation}
\label{eq:normW}
\sum_{\sigma} W(s \to \sigma) = 1, \qquad \forall \sigma.
\end{equation}
In principle, one is interested in finding the whole history of the
system, that is $P(s^0, s^1, \cdots s^t, \cdots)$, given an initial
condition $P(s^0)$. For many purposes, however, a knowledge of the
long--time behaviour is sufficient. In most cases of physical
interest, this can be described by a stationary state $\pi(s)$,
defined as the state which satisfies
\begin{equation}
\label{eq:ss}
\pi(\sigma) = \sum_s \pi(s) W(s \to \sigma)
\end{equation}
(if the system does not reach a stationary state, its
  long--time behaviour may or may not be approximated by suitable
  generalizations of the present approach: for example, for a periodic
  system of period $T$ one would have to consider $T$ time steps at a
  time, which might lead to consider longer--range interactions on the
  graph, making the approach tractable only for small $T$). 
In the
following, we shall comment only briefly on the full problem and
concentrate mainly on finding good approximations to the stationary
state. This is in itself a very difficult (in most cases intractable)
problem, as soon as the transition matrix introduces correlations
between variables at different nodes.

In order to develop approximations, it is useful to consider a
variational formulation of the stationary state problem
\cite{PPM0}. The full history of the system up to a (possibly
infinite) time $t$ can be viewed as the probability distribution which
minimizes the following kinetic generalization of the free energy:
\begin{eqnarray}
  \label{eq:F}
  && {\mathcal F}[P(s^0, s^1, \cdots s^t)] = \sum_{s^0 \cdots s^t} P(s^0,
  s^1, \cdots s^t) \times \nonumber \\
&& \times \left[ 
    -\sum_{\tau = 1}^t \ln W(s^{\tau-1} \to s^\tau) + \ln P(s^0, s^1,
    \cdots s^t) \right]
\end{eqnarray}
subject to the marginalization constraint
\begin{equation}
\label{eq:margt0}
\sum_{s^1 \cdots s^t} P(s^0, s^1, \cdots s^t) = P(s^0)
\end{equation}
(in order to minimize symbol proliferation, in the above equations we
use the same symbol for the argument of ${\mathcal F}$ and the system
history, which actually corresponds to the $\tx{argmin}$ of ${\mathcal
  F}$; moreover, we use the same symbol $P$ for all probability
distributions, and distinguish them only on the basis of their
arguments).

On the other hand, thanks to the Markov property of our kinetics, we
have 
\begin{eqnarray}
\label{eq:history}
&& P(s^0, s^1, \cdots s^t) = P(s^0) \prod_{\tau = 1}^t W(s^{\tau-1} \to
s^\tau) \nonumber \\
&& = P(s^0) \prod_{\tau = 1}^t P(s^\tau \vert s^{\tau-1})
= \frac{\prod_{\tau = 1}^t P(s^{\tau-1},s^\tau)}{\prod_{\tau = 1}^{t-1}
  P(s^\tau)},
\end{eqnarray}
a property similar to the factorization of the equilibrium probability
in one--dimensional Boltzmann problems with nearest--neighbour
interactions, which makes the Bethe approximation exact in such
problems \cite{MyRev}. Our variational functional ${\mathcal F}$ can
then be written as a functional of 2--times and 1--time probabilities
as (for brevity we omit the argument of ${\mathcal F}$)
\begin{eqnarray}
\label{eq:F21}
{\mathcal F} &=& \sum_{\tau = 1}^t \sum_{s^{\tau-1},s^\tau}
P(s^{\tau-1},s^\tau) \left[ - \ln W(s^{\tau-1} \to s^\tau) \right. \nonumber \\
&+& \left. \ln
  P(s^{\tau-1},s^\tau) \right] - \sum_{\tau = 1}^{t-1}
\sum_{s^\tau} P(s^\tau) \ln P(s^\tau),
\end{eqnarray}
to be minimized subject to the marginalization constraints
\begin{eqnarray}
\label{eq:ex-constr1}
P(s^{\tau-1}) &=& \sum_{s^\tau} P(s^{\tau-1},s^\tau), \qquad \tau = 1,
\cdots t \\
\label{eq:ex-constr2}
P(s^\tau) &=& \sum_{s^{\tau-1}} P(s^{\tau-1},s^\tau), \qquad \tau = 1,
\cdots t.
\end{eqnarray}
We could introduce suitable Lagrange multipliers to enforce the
constraints, it is however simpler to use eq.\ \ref{eq:ex-constr1} as
a definition of $P(s^{\tau-1})$. Then, minimizing ${\mathcal F}$ with
respect to the 2--times probabilities we obtain
\begin{equation}
P(s^{\tau-1},s^\tau) = W(s^{\tau-1} \to s^\tau) P(s^{\tau-1}),
\end{equation}
and eq.\ \ref{eq:ex-constr2} reduces to eq.\ \ref{eq:Markov}, showing
that this variational formulation is indeed equivalent to the original
kinetic problem.

In order to obtain a variational formulation for the stationary state
problem it is now sufficient to observe that, assuming that the
long--time kinetics converges to a stationary state, the 1--time and
2--times marginals become time--independent and the density (per unit
time) corresponding to our functional is
\begin{eqnarray}
\label{eq:f21}
f &=& \sum_{s,\sigma}
P(s,\sigma) \left[ - \ln W(s \to \sigma) + \ln
  P(s,\sigma) \right] \nonumber \\ &-& \sum_{s} P(s) \ln P(s),
\end{eqnarray}
to be minimized with the constraint $P(\sigma) = \sum_s
P(s,\sigma)$.

The variational functionals in eqs.\ \ref{eq:F21} and \ref{eq:f21}
(for the stationary state) are used as starting points to develop
variational approximations in the PPM \cite{PPM1,PPM2,PPM3}: for a
given graph $G$, a set $R$ of clusters (subsets of $V$) is selected
according to the principles of the cluster variation method (CVM)
\cite{CVM,An,MyRev}, the 2--times and 1--time entropies are expanded
into a sum of contributions associated to such clusters, and the
resulting functional is minimized with respect to the 1--time and
2--times probability distributions $P(s_\alpha^\tau)$ and
$P(s_\alpha^{\tau-1},s_\alpha^\tau)$ of each cluster $\alpha \in R$,
with the appropriate marginalization constraints. The constrained
variational problem can then be solved by means of simple
generalizations of message--passing algorithms like those developed in
\cite{Yedidia}.
The PPM has been shown, at least in one case \cite{ZM-JSTAT}, to be
equivalent to a technique sometimes called local equilibrium approach,
based on the assumption of a suitable factorization of the stationary
state \cite{Kawasaki,DeLos1,DeLos2}.

Here, however, we would like to take a slightly different route with
respect to PPM, by applying the cluster expansion of the entropy
directly to the functional ${\mathcal F}$ in eq.\ \ref{eq:F}. The
variables $s_i^t$ will be regarded as associated to the nodes, labeled
by the pair $(i,t)$, of an extended graph $G_T$, obtained by time
translation of $G$. 

We shall illustrate the idea with two examples: a simple one, reducing
to a mean--field theory, and a more advanced one, leading to a new and
powerful technique. To fix ideas we shall restrict our discussion to
kinetic Ising--like models with parallel (or synchronous) update,
where the transition matrix has the simple form
\begin{equation}
\label{eq:Wparallel}
W(s \to \sigma) = \prod_{i \in G} W_i(\sigma_i \vert s_{\partial i}).
\end{equation}
In the case of Ising variables, a frequently adopted choice for the
transition matrix is the Glauber one, specified by
\begin{equation}
\label{eq:WGlauber}
W_i(\sigma_i \vert s_{\partial i}) = \frac{\exp[\sigma_i(h_i + \sum_{j
  \in \partial i} J_{ji} s_j)]}{2 \cosh(h_i + \sum_{j
  \in \partial i} J_{ji} s_j)},
\end{equation}
where $h_i$ is a local field, $J_{ji}$ is a coupling (in general
$J_{ji} \ne J_{ij}$) and temperature has been absorbed into fields and
couplings. 

In order to perform a cluster expansion in eq.\ \ref{eq:F}, a key
observation is that if one does not want to introduce additional
approximations, the set $R$ of clusters used in the entropy expansion
should contain the clusters involved in the specification of $W$, in
the present case the star--like clusters $A_{i, t} = \{(i,t)\} \cup
\{(j,t-1), j \in \partial i \}$ with $(i,t) \in G_T$ and $t > 0$. The
simplest possible choice is then to take $A_{i, t}$ as maximal
clusters and expand the entropy term in eq.\ \ref{eq:F} according to
the rules of the CVM \cite{An,MyRev} (we shall call {\em star
  approximation} the resulting method). For simplicity, we shall
consider a locally tree--like graph, without short loops. In such a
case the only intersections of our star clusters are single nodes of
$G_T$, and each node $(i,t)$ appears in $d_i + 1$ star clusters (only
$d_i$ if $t = 0$). The cluster expansion of the entropy in eq.\
\ref{eq:F} is then
\begin{eqnarray}
  \label{eq:mfF}
  {\mathcal F} &\simeq& \sum_{(i,t>0)} \sum_{s_i^t,s_{\partial i}^{t-1}}
  P(s_i^t,s_{\partial i}^{t-1}) \left[ - \ln W_i(s_i^t \vert s_{\partial
      i}^{t-1}) \right. \nonumber \\
  &+& \left. \ln P(s_i^t,s_{\partial i}^{t-1}) \right] -
  \sum_{(i,t>0)} d_i \sum_{s_i^t} P(s_i^t) \ln P(s_i^t) \nonumber 
  \\ &-&
  \sum_i (d_i - 1) \sum_{s_i^0} P(s_i^0) \ln P(s_i^0) 
\end{eqnarray}
where $s_{\partial i}^{t-1} = \{ s_j^{t-1}, j \in \partial i \}$. If
the graph $G$ contains short loops additional terms may enter the
expansion, but the following results can still be used as a low--order
approximation. The above functional must 
be minimized with respect to the star cluster and single node
probability distributions, subject to the marginalization constraints
\begin{eqnarray}
\label{eq:star-node1}
&& P(s_j^{t-1}) = \sum_{s_i^t, s_{\partial i \setminus j}^{t-1}}
P(s_i^t,s_{\partial i}^{t-1}), \\
\label{eq:star-node2}
&& P(s_i^t) = \sum_{s_{\partial i}^{t-1}} P(s_i^t,s_{\partial i}^{t-1}).
\end{eqnarray}
Using eq.\ \ref{eq:star-node1} as a definition for the
single--node probabilities and minimizing 
${\mathcal F}$ in eq.\ \ref{eq:mfF} with respect to the star cluster
probabilities we obtain
\begin{equation}
P(s_i^t,s_{\partial i}^{t-1}) = W_i(s_i^t \vert s_{\partial i}^{t-1})
\prod_{j \in \partial i} P(s_j^{t-1}),
\end{equation}
while eq.\ \ref{eq:star-node2} becomes
\begin{equation}
P(s_i^t) = \sum_{s_{\partial i}^{t-1}} W_i(s_i^t \vert s_{\partial i}^{t-1})
\prod_{j \in \partial i} P(s_j^{t-1}),
\end{equation}
which in the stationary limit reduces to
\begin{equation}
P(\sigma_i) = \sum_{s_{\partial i}} W_i(\sigma_i \vert s_{\partial i})
\prod_{j \in \partial i} P(s_j),
\end{equation}
and can be used as a basis for an iterative solution. Our star
approximation is then a mean--field--like approximation, structurally
similar to the hard--spin mean--field theory \cite{Berker,Maritan} for
the equilibrium problem, where the stationary state is assumed to
factor into a product of single node probabilities, as discussed for
example in \cite{DeLos1,DeLos2}.

In order to go beyond this mean--field approximation, one should at
least take into account correlations in $s_{\partial i}^{t-1}$. If the
graph $G$ does not contain short loops, these correlations will be
primarily due to the interactions that variables in $s_{\partial
  i}^{t-1}$ have with $s_i^{t-2}$. This observation naturally leads to
introduce a new approximation, by choosing as maximal clusters in our
entropy expansion the diamond--like clusters $B_{i, t} = \{(i,t)\}
\cup \{(j,t-1), j \in \partial i \} \cup \{(i,t-2)\}$ with $(i,t) \in
G_T$ and $t > 1$ (we shall call {\em diamond approximation} the
resulting method). The choice of these clusters, besides being quite
natural, is inspired by the dynamic cavity method, whose recursive
equations involve the same sets of variables \cite{DynCavity3}, but
the resulting method will be more general and more powerful. In a
graph without short loops, the cluster expansion \cite{An,MyRev} of
eq.\ \ref{eq:F} based on our diamond--like clusters contains also
terms corresponding to the following clusters: the pairs $\{
(i,t),(j,t-1) \}$, with $j \in \partial i$ (whose entropy terms will
have a coefficient -1, since they are subclusters of 2 different
diamond clusters); the single nodes $(i,t)$ (with coefficient $d_i -
1$, since they are subclusters of $d_i + 2$ diamond clusters and $2
d_i$ pair clusters). We obtain
\begin{eqnarray}
  \label{eq:diamondF}
  {\mathcal F} &\simeq& \sum_{(i,t)} \sum_{s_i^t,s_{\partial i}^{t-1},s_i^{t-2}}
  P(s_i^t,s_{\partial i}^{t-1},s_i^{t-2}) \times \nonumber \\
&\times& \left[ - \ln W_i(s_i^t \vert s_{\partial
      i}^{t-1}) + \ln P(s_i^t,s_{\partial i}^{t-1},s_i^{t-2}) \right]
  \nonumber \\ &-&
  \sum_{(i,t-2)} \sum_{j \in \partial i} \sum_{s_j^{t-1},s_i^{t-2}}
  P(s_j^{t-1},s_i^{t-2}) \ln P(s_j^{t-1},s_i^{t-2}) \nonumber \\
&+&
  \sum_{(i,t-2)} (d_i - 1) \sum_{s_i^{t-2}} P(s_i^{t-2}) \ln
  P(s_i^{t-2}) \nonumber \\ &+& \tx{boundary terms}
\end{eqnarray}
(there is no need to specify boundary terms since our main interest is
the stationary state), with the following pair--node
\begin{eqnarray}
\label{eq:pair-node1}
P(s_i^{t-2}) &=& \sum_{s_j^{t-1}} P(s_j^{t-1},s_i^{t-2}) \\
\label{eq:pair-node2}
P(s_j^{t-1}) &=&\sum_{s_i^{t-2}} P(s_j^{t-1},s_i^{t-2}) 
\end{eqnarray}
and diamond--pair 
\begin{eqnarray}
\label{eq:diamond-pair1}
P(s_j^{t-1},s_i^{t-2}) &=& \sum_{s_i^t,s_{\partial i \setminus
    j}^{t-1}} P(s_i^t,s_{\partial i}^{t-1},s_i^{t-2}) \\
\label{eq:diamond-pair2}
P(s_i^t,s_j^{t-1}) &=& \sum_{s_{\partial i \setminus
    j}^{t-1},s_i^{t-2}} P(s_i^t,s_{\partial i}^{t-1},s_i^{t-2})
\end{eqnarray}
marginalization constraints. Proceeding as before, we define
$P(s_i^{t-2})$ and $P(s_j^{t-1},s_i^{t-2})$ as marginals of the
diamond cluster probabilities using eqs.\ \ref{eq:pair-node1} and
\ref{eq:diamond-pair1}. Minimizing ${\mathcal F}$ in eq.\
\ref{eq:diamondF} with respect to the diamond cluster probabilities we
then obtain
\begin{eqnarray}
\label{eq:pdiamond}
&& P(s_i^t,s_{\partial i}^{t-1},s_i^{t-2}) = W_i(s_i^t \vert s_{\partial
      i}^{t-1}) \times \nonumber \\
&& \times 
 [P(s_i^{t-2})]^{1-d_i} \prod_{j \in \partial i} P(s_j^{t-1},s_i^{t-2}), 
\end{eqnarray}
or equivalently, in terms of conditional pair probabilities,
\begin{equation}
\label{eq:pdiamondcond}
P(s_i^t,s_{\partial i}^{t-1},s_i^{t-2}) = W_i(s_i^t \vert s_{\partial
      i}^{t-1})  
    P(s_i^{t-2}) \prod_{j \in \partial i} P(s_j^{t-1} \vert
      s_i^{t-2}). 
\end{equation}

Eq.\ \ref{eq:pdiamond} (or \ref{eq:pdiamondcond}), together with eqs.\
\ref{eq:pair-node2} and \ref{eq:diamond-pair2}, provides the solution
to our problem. Given the single--node probabilities at time $t-2$,
and the pair probabilities at times $(t-2,t-1)$, we can use these
equations to find the same probabilities one time step later. We have
therefore an iterative scheme which (if convergent) provides, in the
long time limit, an approximation to the stationary state. 

The accuracy of the method will be tested numerically, in specific
cases, in the next section. Here it is interesting to qualitatively
compare our approximation with the recently proposed dynamic cavity
method
\cite{DynCavity00,DynCavity0,DynCavity1,DynCavity2,DynCavity3}. The
latter method, which needs the so--called time--factorization (or
one--time) approximation \cite{DynCavity1,DynCavity2,DynCavity3} to
make it tractable, is a message--passing algorithm where messages
(which can be thought of as a suitable parametrization of cavity
marginals) are exchanged between neighbouring nodes. In order to
understand the difference between our method and dynamic cavity, it is
useful to rewrite our eq.\ \ref{eq:pair-node2}, using eqs.\
\ref{eq:diamond-pair2} and \ref{eq:pdiamondcond}, with the following
result:
\begin{equation}
\label{eq:recpnode}
P(s_i^t) = \sum_{s_{\partial i}^{t-1},s_i^{t-2}}
W_i(s_i^t \vert s_{\partial
  i}^{t-1}) 
P(s_i^{t-2}) \prod_{j \in \partial i} P(s_j^{t-1} \vert s_i^{t-2}). 
\end{equation}
This is structurally similar to the recursive equation for single node
marginals found in \cite{DynCavity2}. The latter however contains, in
place of our conditional pair probabilities, certain quantities
(messages) which in the case of Ising variables are parametrized as
\begin{equation}
\label{eq:DynCavity}
\mu(s_j^{t-1} \vert s_i^{t-2}) = \frac
{\exp[s_j^{t-1}(u_{j \to i} + J_{ij} s_i^{t-2})]}
{2 \cosh(u_{j \to i} + J_{ij} s_i^{t-2})},
\end{equation}
where $u_{j \to i}$ is determined recursively. It has to be noticed
here that our conditional pair probabilities can be written in a
similar form, but the corresponding effective interaction between
$s_i^{t-2}$ and $s_j^{t-1}$ is not constrained to $J_{ij}$. In a
general problem it can take any value, although it reduces to the true
coupling $J_{ij}$ in the fully symmetric case ($J_{ji} = J_{ij},
\forall (i,j) \in E$), when the two methods become equivalent (and
exact if $G$ is a tree): 
see the Appendix for a proof of
  this equivalence. 
As a consequence, our method has more parameters
than dynamic cavity, and needs more equations. Indeed we have to solve
equations for both single--node and pair probabilities, while dynamic
cavity with the time--factorization approximation is expressed through
recursive equations on single--node (cavity) marginals only
\cite{DynCavity3}. One might expect this to affect the relative
performance of the two methods, making ours slower by a factor of $q$
for $q$--state variables, but in the next section we shall see that
this is not the case.

\section{Results}

Here we compare the star-- and diamond--cluster approximations we have
derived in the previous section with approximations of comparable
complexity: the dynamic cavity method in the time--factorization
approximation \cite{DynCavity2,DynCavity3}; a 3--times mean--field
approximation proposed in \cite{DeLos2}, which involves a summation
over the second neighbourhood of a node; the so--called naive mean
field (see e.g.\ \cite{DynCavity3}). We also considered the pair
approximation \cite{Matsuda,DeLos1}, based on assuming a factorization
of the stationary state probability at the level of equal--time
neighbouring pairs, however we do not report the corresponding results
since they are indistinguishable from those of our star approximation
on the scale of the graphs. For reference, exact or Monte Carlo
results will be used, depending on the size of the graph. We shall
consider Ising--like models, with the transition matrices defined in
eqs.\ \ref{eq:Wparallel} and \ref{eq:WGlauber}, with random,
independently drawn, fields and couplings. $h_i$ will be taken from a
uniform distribution in $(-1/2,1/2)$, while $J_{ij} \ne J_{ji}$ will
be taken from a uniform distribution in $(-J_0,J_0)$. The quantity
$\delta m = \sqrt{\frac{1}{N} (m_i - m_i^{\tx{exact}})^2}$ is used as
a measure of the performance of the methods, where $m_i$ is the
estimate of $\langle s_i \rangle$ in the stationary state provided by
an approximate method and $m_i^{\tx{exact}}$ is the corresponding
exact result.

\begin{figure}
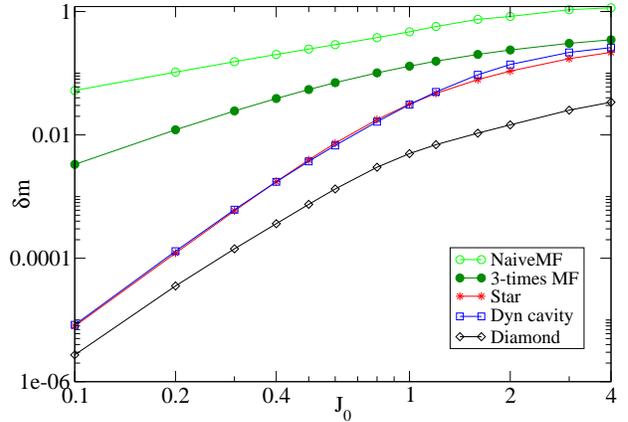

\onefigure[width=0.45\textwidth]{Fig1a.eps}
\caption{$\delta m$ vs $J_0$ in a random graph with $N = 14$ nodes for
  various approximations: naive mean--field (light green, open
  circles), 3--times mean--field (dark green, solid circles), our star
  approximation (red, stars), dynamic cavity (blue, squares) and our
  diamond approximation (black, diamonds).}
\label{fig:small}
\end{figure}

In fig.\ \ref{fig:small} a random graph of regular degree 3 ($d_i = 3,
\forall i \in G$) and $N = 14$ nodes was considered, and $\delta m$ is
reported as a function of $J_0$ for the various methods. It is seen
that the star approximation outperforms the other mean--field
techniques and is practically equivalent to the dynamic cavity method,
while the diamond approximation outperforms (by almost an order of
magnitude for large $J_0$) also the dynamic cavity method.

\begin{figure}
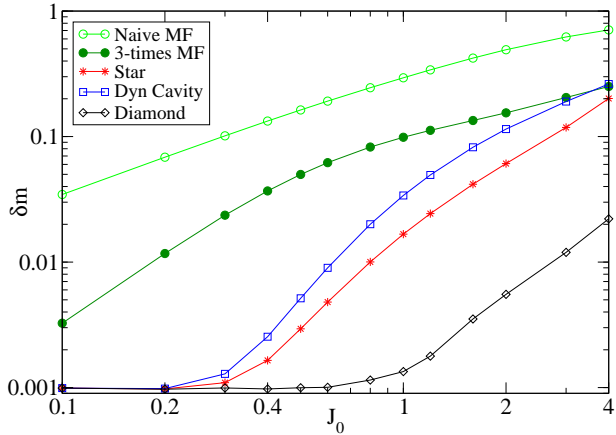

\onefigure[width=0.45\textwidth]{Fig1b.eps}
\caption{Same as fig.\ \protect\ref{fig:small} for a graph with $N =
  1000$ nodes.}
\label{fig:large}
\end{figure}

Fig.\ \ref{fig:large} contains a similar plot in the case of a graph
with $N = 10^3$ nodes, and Monte Carlo results were used in place of
exact ones. For each data point, Monte Carlo results are obtained by
averaging over $10^6$ time steps, after waiting $10^5$ time steps for
reaching the stationary state. The relative performance of the various
methods is the same as in the case of the small graph. The plateau in
the bottom--left part of the figure means that for small enough $J_0$
some approximations are more accurate than the Monte Carlo
simulations. 

\begin{figure}
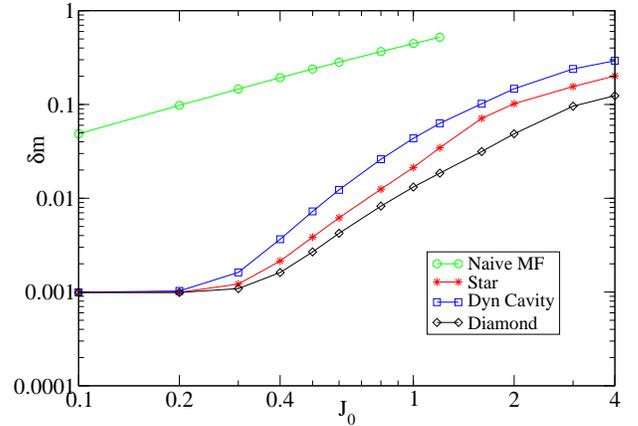

\onefigure[width=0.45\textwidth]{Fig3.eps}
  \caption{Same as fig.\ \protect\ref{fig:small} for a square lattice
    with $N = 30^2$ nodes.}
\label{fig:square}
\end{figure}

As a further check, we considered a square lattice with $N = 30^2$
nodes and periodic boundary conditions. Corresponding results, using
the same parameters as in previous cases, are reported in Fig.\
\ref{fig:square}. Here, the 3--times mean--field approximation was not
used, since it becomes too slow (much slower than Monte Carlo
simulations) due to the sum over the second neighbourhood. Moreover,
the naive mean--field theory does not converge for $J_0 > 1.2$. The
performance of the other approximations is reduced, but the general
picture is qualitatively the same as before. As far as the diamond
approximation is concerned, we must stress that here we have used
eqs.\ \ref{eq:pdiamond}, \ref{eq:pair-node2} and
\ref{eq:diamond-pair2}, which were developed in the previous section
for a graph without short loops. Taking into account short loops in
the entropy expansion should result in a more accurate (although
slower) algorithm.

\begin{figure}
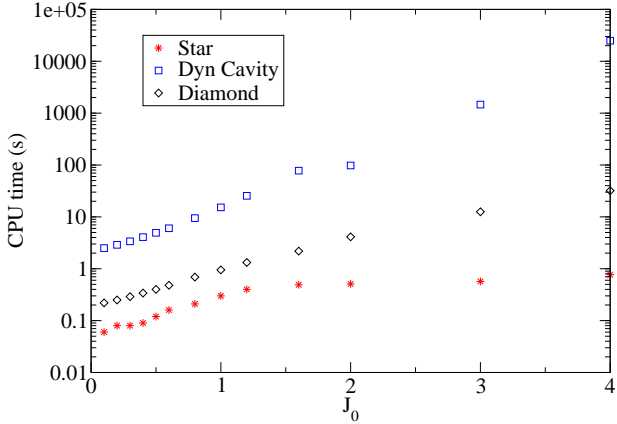

\onefigure[width=0.45\textwidth]{Fig3-Times.eps}
  \caption{CPU times, evaluated on a 1.4 GHz 64-bit processor, for the
  case of Fig.\ \protect\ref{fig:square}. Naive MF results are too
  small ($\sim 10^{-2}$ s) to provide a reliable estimate.}
\label{fig:square-times}
\end{figure}

It is also interesting to observe that the star and
diamond approximations are much faster than the dynamic cavity method,
see Fig.\ \ref{fig:square-times}. At $J_0 = 0.1$ the diamond approximation
is 12 times faster than dynamic cavity, and this figure increases with
$J_0$. Since a single iteration of the dynamic cavity is
computationally comparable to a single iteration of the diamond
approximation (and becomes simpler for large $q$), this means that the
latter requires a smaller number of iterations.

\begin{figure}
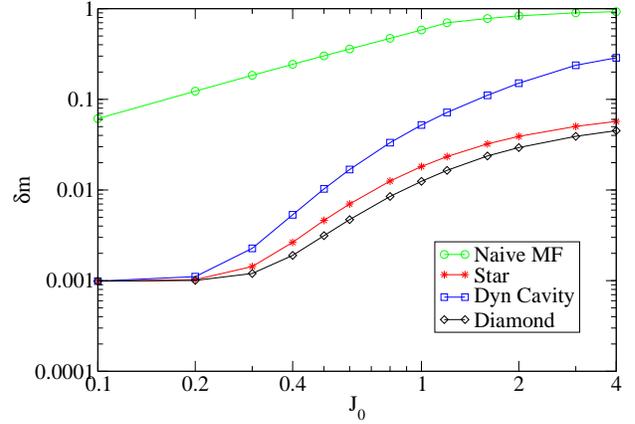

\onefigure[width=0.45\textwidth]{FigCubic.eps}
  \caption{Same as fig.\ \protect\ref{fig:small} for a cubic lattice
    with $N = 10^3$ nodes.}
\label{fig:cubic}
\end{figure}

Finally, in order to further explore the behaviour of our approach in
the case of a graph with many short loops, we repeated the above tests
for a simple cubic lattice with $N = 10^3$ nodes and periodic boundary
conditions. Results are reported in Figs.\ \ref{fig:cubic} ($\delta
m$) and \ref{fig:cubic-times} (CPU times). 

The picture is similar to the square lattice case: naive mean--field
is unreliable, while the other approximations perform better than
Monte Carlo at $J_0 \le 0.2$ and show an error which increases with
$J_0$, with the diamond approximation exhibiting the smallest $\delta
m$, followed by the star approximation and the dynamic cavity. Again,
the diamond approximation is faster than dynamic cavity, more
precisely 20 times faster at $J_0 = 0.1$, with this ratio increasing
with $J_0$. In particular, at $J_0 = 4$, the dynamic cavity seems to
converge, albeit in a prohibitively long time: the reported value of
$\delta m$ was obtained after $10^6$ iterations, and the CPU time to
convergence was extrapolated.

\begin{figure}
\onefigure[width=0.45\textwidth]{FigCubic-Times.eps}
  \caption{Same as fig.\ \protect\ref{fig:square-times} for a cubic lattice
    with $N = 10^3$ nodes.}
\label{fig:cubic-times}
\end{figure}

\section{Discussion}

We have introduced a new variational approach to the stationary states
of kinetic Ising--like models. The approximation is based on the
cluster expansion of the entropy term appearing in a variational
functional which is minimized by the system history. The new feature
is that the cluster expansion is performed on the full functional
(eq.\ \ref{eq:F}) and not on a 2--times functional (eqs.\ \ref{eq:F21}
and \ref{eq:f21}) as in the path probability method. This leads us to
use as main clusters subsets of nodes which are more general than
those obtained by simple time translation of the typical clusters used
in equilibrium approximations. The approximation, in its present
formulation, is limited to discrete time kinetics. 

We have illustrated our idea with two examples. At the lowest level we
have obtained a well--known mean--field theory, here called the star
approximation. Adding only one node to the main clusters we have
obtained a new method, here called diamond approximation. This was
tested against other methods of comparable computational complexity on
models defined on random graphs and on graphs with many short loops (a
square lattice and a simple cubic lattice) with random fields and
random interactions of varying strength. The diamond approximation
turned out to be more accurate and faster than other methods of
similar complexity, including the ordinary pair approximation and the
recently proposed dynamic cavity method.

These results suggest that several extensions and improvements might
be worth considering. 
In the case of graphs with many short loops, like Euclidean lattices
in 2 or 3 dimensions, one could try to get further improvements by
introducing  
additional terms in the entropy expansion. In this case,
message--passing algorithms like the generalized belief propagation
\cite{Yedidia} may be needed for the minimization. In case of
convergence problems, provably convergent algorithms \cite{HAK,MyRev}
may be used instead. Moreover, the use of larger clusters can be
considered. The simplest example in this direction is including node
$(i,t-1)$ in the diamond--like cluster $B_{i,t}$: this will allow to
introduce self--interaction terms like $J_{ii} s_i^{t-1} s_i^t$ in the
transition matrix, which is relevant e.g.\ for models of epidemic
processes. Another line of investigation could involve the use of
different kinds of transition matrices, the simplest examples being
asynchronous updates and exchange dynamics: the latter, in particular,
may need the introduction of larger clusters. A further important
question to address is related to the applicability of these methods
to an approximate description of the transient, although it was shown
in \cite{DynCavity1,DynCavity3} that these mean--field like techniques
are more appropriate for the stationary state. 
Work is in
progress along these lines. 

\section{Appendix}


  In the case of symmetric interactions ($J_{ij} = J_{ji}, \forall
  (i,j) \in E$), it was shown in \cite{DynCavity2} that a fixed point
  of ordinary belief propagation (BP) corresponds to a stationary
  state of the dynamic cavity method in the time--factorization
  approximation. In this Appendix we shall show that this property is
  shared by our diamond approximation, thereby establishing an
  equivalence between the dynamic cavity method and the diamond
  approximation. As shown in the previous sections, for non--symmetric
  interactions the two methods give different results, so the
  equivalence is limited to the case of symmetric interactions.

  Let us consider the stationary state problem defined by the
  transition matrix eq.\ \ref{eq:WGlauber}, with $J_{ij} = J_{ji}$,
  and the corresponding equilibrium problem, defined by the Ising
  Hamiltonian 
  \begin{equation}
    H = - \sum_{i \in G} h_i s_i - \sum_{(i,j) \in E} J_{ij} s_i s_j.
  \end{equation}
  In the following, we will use a symmetrized form of the above
  Hamiltonian, where $H = \displaystyle{\sum_{(i,j) \in E}}
  H_{ij}(s_i,s_j)$ and
  \begin{equation}
    H_{ij}(s_i,s_j) = - J_{ij} s_i s_j - \frac{h_i}{d_i} s_i -
    \frac{h_j}{d_j} s_j. 
  \end{equation}
  Moreover, we shall define $\psi_{ij}(s_i,s_j) = \exp\left[ -
    H_{ij}(s_i,s_j) \right]$.

  BP \cite{Mezard,MyRev} provides an approximate
  solution to the above equilibrium problem, with the single--node and
  pair marginals given by
  \begin{eqnarray}
    P(s_i) &=& \frac{1}{Z_i} \prod_{k \in \partial i} m_{k \to i}(s_i)
    , \label{eq:PnodeBP} \\
    P(s_i,s_j) &=& \frac{\psi_{ij}(s_i,s_j)}{Z_{ij}} 
    \prod_{k \in \partial i \setminus j} m_{k \to i}(s_i) \times
    \nonumber \\
    && \prod_{l \in \partial j \setminus i} m_{l \to j}(s_j). \label{eq:PpairBP} 
  \end{eqnarray}
  In the above equations, $m_{k \to i}(s_i)$ is called the message
  from node $k$ to node $i$, while $Z_i$ and $Z_{ij}$ are
  normalization constants. Messages are determined by imposing the
  marginalization constraints $P(s_i) = \sum_{s_j} P(s_i,s_j)$, which
  yield (up to normalization)
  \begin{equation}
    m_{j \to i}(s_i) \propto \sum_{s_j} \psi_{ij}(s_i,s_j)
    \prod_{l \in \partial j \setminus i} m_{l \to j}(s_j),
    \label{eq:BP}
  \end{equation}
  usually solved by iteration to a fixed point. 

  We can now show that a fixed point of the BP
  equations corresponds to a stationary state of our diamond
  approximation. In eqs.\ \ref{eq:pdiamond} and \ref{eq:pdiamondcond},
  let us suppose that the transition matrix is given by eq.\
  \ref{eq:WGlauber}, and rewrite it as 
  \begin{equation}
    W(s_i^t \vert s_{\partial i}^{t-1}) = \frac
    {\prod_{k \in \partial i} \psi_{ki}(s_k^{t-1},s_i^t)}
    {\sum_{s'} \prod_{k \in \partial i} \psi_{ki}(s_k^{t-1},s')}.
  \end{equation}
  Let us also suppose that $P(s_i^{t-2})$ and $P(s_j^{t-1},s_i^{t-2})$
  have the BP form, eqs.\ \ref{eq:PnodeBP} and
  \ref{eq:PpairBP} respectively. We then obtain
  \begin{eqnarray}
    && P(s_i^t,s_{\partial i}^{t-1},s_i^{t-2}) = W_i(s_i^t \vert s_{\partial
      i}^{t-1}) \times \nonumber \\
    && \prod_{k \in \partial i} \left[ \psi_{ik}(s_i^{t-2},s_k^{t-1})
      \prod_{l \in \partial k \setminus i} m_{l \to k}(s_k^{t-1}) \right].
  \end{eqnarray}
  We can now use eq.\ \ref{eq:diamond-pair2} and obtain
  \begin{eqnarray}
    && P(s_i^t,s_j^{t-1}) = \sum_{s_{\partial i \setminus
        j}^{t-1}} W_i(s_i^t \vert s_{\partial
      i}^{t-1}) \prod_{k \in \partial i} \prod_{l \in \partial k
      \setminus i} m_{l \to k}(s_k^{t-1}) \times \nonumber \\
    && \sum_{s_i^{t-2}} \prod_{k \in \partial i}
    \psi_{ki}(s_k^{t-1},s_i^{t-2}), 
  \end{eqnarray}
  where the last sum cancels the denominator in the transition matrix
  thanks to the symmetry property of the interactions, yielding (up to
  normalization) 
  \begin{eqnarray}
    && P(s_i^t,s_j^{t-1}) \propto \nonumber \\
    && \sum_{s_{\partial i \setminus j}^{t-1}} \prod_{k \in \partial i} 
    \left[ \psi_{ki}(s_k^{t-1},s_i^t)
      \prod_{l \in \partial k \setminus i} m_{l \to k}(s_k^{t-1})
    \right] = \nonumber \\
    && \psi_{ji}(s_j^{t-1},s_i^t)
      \prod_{l \in \partial j \setminus i} m_{l \to j}(s_j^{t-1})
      \times \nonumber \\
      && \prod_{k \in \partial i \setminus j} \sum_{s_k^{t-1}} 
      \left[ \psi_{ki}(s_k^{t-1},s_i^t)
        \prod_{l \in \partial k \setminus i} m_{l \to k}(s_k^{t-1})
    \right]. 
  \end{eqnarray}
  Eventually, using eq.\ \ref{eq:BP}, we obtain
  \begin{eqnarray}
    P(s_i^t,s_j^{t-1}) &\propto& \psi_{ji}(s_j^{t-1},s_i^t)
      \prod_{l \in \partial j \setminus i} m_{l \to j}(s_j^{t-1})
      \times \nonumber \\
      && \prod_{k \in \partial i \setminus j} m_{k \to i}(s_i^{t-1}),
  \end{eqnarray}
  which shows that the BP fixed point is also a fixed point of the
  diamond approximation. 

\end{document}